\documentstyle[twoside,psfig]{article}
\def\vev#1{{\langle#1\rangle}}
\newcommand{\pa}{a^{\prime}}
\newcommand{\pn}{n^{\prime}}
\newcommand{\fpa}{\frac{\pa}{a}}
\newcommand{\fpn}{\frac{\pn}{n}}
\catcode`\@=11
\long\def\@makefntext#1{
\protect\noindent \hbox to 3.2pt {\hskip-.9pt  
$^{{\eightrm\@thefnmark}}$\hfil}#1\hfill}		

\def\@makefnmark{\hbox to 0pt{$^{\@thefnmark}$\hss}}	
	
\def\ps@myheadings{\let\@mkboth\@gobbletwo
\def\@oddhead{\hbox{}
\rightmark\hfil\eightrm\thepage}   
\def\@oddfoot{}\def\@evenhead{\eightrm\thepage\hfil
\leftmark\hbox{}}\def\@evenfoot{}
\def\sectionmark##1{}\def\subsectionmark##1{}}

\oddsidemargin=\evensidemargin
\newcounter{sectionc}\newcounter{subsectionc}\newcounter{subsubsectionc}
\renewcommand{\section}[1] {\vspace{12pt}\addtocounter{sectionc}{1} 
\setcounter{subsectionc}{0}\setcounter{subsubsectionc}{0}\noindent 
	{\tenbf\thesectionc. #1}\par\vspace{5pt}}
\renewcommand{\subsection}[1] {\vspace{12pt}\addtocounter{subsectionc}{1} 
	\setcounter{subsubsectionc}{0}\noindent 
	{\bf\thesectionc.\thesubsectionc. {\kern1pt \bfit #1}}\par\vspace{5pt}}
\renewcommand{\subsubsection}[1] {\vspace{12pt}\addtocounter{subsubsectionc}{1}
	\noindent{\tenrm\thesectionc.\thesubsectionc.\thesubsubsectionc.
	{\kern1pt \tenit #1}}\par\vspace{5pt}}

\newcounter{appendixc}
\newcounter{subappendixc}[appendixc]
\newcounter{subsubappendixc}[subappendixc]
\renewcommand{\thesubappendixc}{\Alph{appendixc}.\arabic{subappendixc}}
\renewcommand{\thesubsubappendixc}
	{\Alph{appendixc}.\arabic{subappendixc}.\arabic{subsubappendixc}}

\renewcommand{\appendix}[1] {\vspace{12pt}
        \refstepcounter{appendixc}
        \setcounter{figure}{0}
        \setcounter{table}{0}
        \setcounter{lemma}{0}
        \setcounter{theorem}{0}
        \setcounter{corollary}{0}
        \setcounter{definition}{0}
        \setcounter{equation}{0}
        \renewcommand{\thefigure}{\Alph{appendixc}.\arabic{figure}}
        \renewcommand{\thetable}{\Alph{appendixc}.\arabic{table}}
        \renewcommand{\theappendixc}{\Alph{appendixc}}
        \renewcommand{\thelemma}{\Alph{appendixc}.\arabic{lemma}}
        \renewcommand{\thetheorem}{\Alph{appendixc}.\arabic{theorem}}
        \renewcommand{\thedefinition}{\Alph{appendixc}.\arabic{definition}}
        \renewcommand{\thecorollary}{\Alph{appendixc}.\arabic{corollary}}
        \renewcommand{\theequation}{\Alph{appendixc}.\arabic{equation}}
        \noindent{\tenbf Appendix \theappendixc #1}\par\vspace{5pt}}
\newcommand{\subappendix}[1] {\vspace{12pt}
        \refstepcounter{subappendixc}
        \noindent{\bf Appendix \thesubappendixc. {\kern1pt \bfit #1}}
	\par\vspace{5pt}}
\newcommand{\subsubappendix}[1] {\vspace{12pt}
        \refstepcounter{subsubappendixc}
        \noindent{\rm Appendix \thesubsubappendixc. {\kern1pt \tenit #1}}
	\par\vspace{5pt}}

\newcommand{\textlineskip}{\baselineskip=13pt}
\newcommand{\smalllineskip}{\baselineskip=10pt}

\def\eightcirc{
\begin{picture}(0,0)
\put(4.4,1.8){\circle{6.5}}
\end{picture}}
\def\eightcopyright{\eightcirc\kern2.7pt\hbox{\eightrm c}} 

\newcommand{\copyrightheading}[1]
	{\vspace*{-2.5cm}\smalllineskip{\flushleft
	{\footnotesize International Journal of Modern Physics A, #1}\\
	{\footnotesize $\eightcopyright$\, World Scientific Publishing
	 Company}\\
	 }}

\def\abstracts#1#2#3{{
	\centering{\begin{minipage}{4.5in}\baselineskip=10pt\footnotesize
	\parindent=0pt #1\par 
	\parindent=15pt #2\par
	\parindent=15pt #3
	\end{minipage}}\par}} 

\newcommand{\bibit}{\nineit}

\renewenvironment{thebibliography}[1]
	{\frenchspacing
	 \ninerm\baselineskip=11pt
	 \begin{list}{\arabic{enumi}.}
	{\usecounter{enumi}\setlength{\parsep}{0pt}
	 \setlength{\leftmargin 12.7pt}{\rightmargin 0pt} 
	 \setlength{\itemsep}{0pt} \settowidth
	{\labelwidth}{#1.}\sloppy}}{\end{list}}

\newcounter{itemlistc}
\newcounter{romanlistc}
\newcounter{alphlistc}
\newcounter{arabiclistc}

\newcommand{\fcaption}[1]{
        \refstepcounter{figure}
        \setbox\@tempboxa = \hbox{\footnotesize Fig.~\thefigure. #1}
        \ifdim \wd\@tempboxa > 5in
           {\begin{center}
        \parbox{5in}{\footnotesize\smalllineskip Fig.~\thefigure. #1}
            \end{center}}
        \else
             {\begin{center}
             {\footnotesize Fig.~\thefigure. #1}
              \end{center}}
        \fi}

\newcommand{\tcaption}[1]{
        \refstepcounter{table}
        \setbox\@tempboxa = \hbox{\footnotesize Table~\thetable. #1}
        \ifdim \wd\@tempboxa > 5in
           {\begin{center}
        \parbox{5in}{\footnotesize\smalllineskip Table~\thetable. #1}
            \end{center}}
        \else
             {\begin{center}
             {\footnotesize Table~\thetable. #1}
              \end{center}}
        \fi}

\def\@citex[#1]#2{\if@filesw\immediate\write\@auxout
	{\string\citation{#2}}\fi
\def\@citea{}\@cite{\@for\@citeb:=#2\do
	{\@citea\def\@citea{,}\@ifundefined
	{b@\@citeb}{{\bf ?}\@warning
	{Citation `\@citeb' on page \thepage \space undefined}}
	{\csname b@\@citeb\endcsname}}}{#1}}

\newif\if@cghi
\def\cite{\@cghitrue\@ifnextchar [{\@tempswatrue
	\@citex}{\@tempswafalse\@citex[]}}
\def\citelow{\@cghifalse\@ifnextchar [{\@tempswatrue
	\@citex}{\@tempswafalse\@citex[]}}
\def\@cite#1#2{{$\null^{#1}$\if@tempswa\typeout
	{IJCGA warning: optional citation argument 
	ignored: `#2'} \fi}}

\def\pmb#1{\setbox0=\hbox{#1}
	\kern-.025em\copy0\kern-\wd0
	\kern.05em\copy0\kern-\wd0
	\kern-.025em\raise.0433em\box0}

\def\fnt#1#2{\footnotetext{\kern-.3em
	{$^{\mbox{\scriptsize #1}}$}{#2}}}

\def\fpage#1{\begingroup
\voffset=.3in
\thispagestyle{empty}\begin{table}[b]\centerline{\footnotesize #1}
	\end{table}\endgroup}

\def\runninghead#1#2{\pagestyle{myheadings}
\markboth{{\protect\footnotesize\it{\quad #1}}\hfill}
{\hfill{\protect\footnotesize\it{#2\quad}}}}
\font\tenrm=cmr10
\font\tenit=cmti10 
\font\tenbf=cmbx10
\font\bfit=cmbxti10 at 10pt
\font\ninerm=cmr9
\font\nineit=cmti9

\font\eightrm=cmr8

\def\qed{\hbox{${\vcenter{\vbox{			
   \hrule height 0.4pt\hbox{\vrule width 0.4pt height 6pt
   \kern5pt\vrule width 0.4pt}\hrule height 0.4pt}}}$}}

\begin{document}

\runninghead{Cosmology and Hierarchy in Stabilized Randall-Sundrum Models}{Cosmology and Hierarchy in Stabilized Randall-Sundrum Models}

\normalsize\textlineskip
\thispagestyle{empty}
\setcounter{page}{1}

\copyrightheading{}			

\vspace*{0.88truein}

\fpage{1}
\centerline{\bf COSMOLOGY AND HIERARCHY IN STABILIZED}
\vspace*{0.035truein}
\centerline{\bf RANDALL-SUNDRUM MODELS}
\vspace*{0.37truein}
\centerline{\footnotesize DANNY MARFATIA}
\vspace*{0.015truein}
\centerline{\footnotesize\it Department of Physics, University of Wisconsin--Madison, WI 53706}
\vspace*{0.21truein}
\abstracts{We consider the cosmology and hierarchy of scales 
in models with branes immersed in 
a five-dimensional curved spacetime subject to radion stabilization. 
The universe naturally find itself in the 
radiation-dominated epoch when the inter-brane spacing is static and stable, 
 independent of the form of the stabilizing potential. We recover the 
standard Friedmann equations without assuming 
a specific form for the bulk energy-momentum tensor. We address the 
hierarchy problem in the context of a quartic and exponential stabilizing
potential, and find that in either case the presence of a negative
tension brane is required and that the
string scale can be as low as the electroweak scale. 
In the situation of self-tuning branes (corresponding to an 
exponential potential) 
where the bulk cosmological constant is set to zero, 
the brane tensions have hierarchical values.}{}{}

\textlineskip			
\vspace*{12pt}			

We summarize the results of Ref.$^1$ We assume the presence of three 
3-branes in the space $(-\infty,\infty)$, 
 at $y_0=0\,$, $y_1$ and $y_2$ with the observable brane
 located at $y_0\,$. We will refer to the branes at $y_1$ and $y_2$ as
 hidden branes.  The four-dimensional metric on
the brane labelled by its position $y_i$ is 
$g_{\mu \nu}^{(i)} (x^{\mu}) \equiv g_{\mu \nu}(x^{\mu}, y=y_i)\,$,
where $g_{AB}$ is the five-dimensional metric, $A,B=0,1,2,3,5$ and 
$\mu,\nu=0,1,2,3$. The five-dimensional Einstein equations
including a radion field $\phi$ are,
$G_{AB}=\kappa^2\,T_{AB}= 
{1 \over 2 }\, ( \partial_A \phi\,  \partial_B \phi
-{1 \over 2 }\, g_{AB}\, ( \partial \phi)^2)-
\kappa^2\, g_{AB} \Lambda (\phi) -\kappa^2\, \sum_{i=0}^{2}
V_i (\phi) \sqrt{{g^{(i)} \over g}} ~g_{\mu \nu}^{(i)} 
~\delta^\mu_A \delta^\nu_B ~\delta(y-y_i)\,,$
where $\kappa^2=8\,\pi\,G_N^{(5)}=M_X^{-3}$ is the 
five-dimensional coupling constant of gravity and $M_X$ is the Planck scale 
in five dimensions. $V_i(\phi (y_i))$ is the tension of the brane
at $y_i$ and $\Lambda(\phi)$ is the potential of the radion in the bulk
and is interpreted as the cosmological constant although it has 
a $\phi$-dependence. We allow it to be discontinuous at the branes, but continuous in each section.
 We write $\Lambda(\phi)$ as $\Lambda_0(\phi)\,$ if $\,y<0\,$, $\Lambda_1(\phi)\,$ if $\,0<y<y_1\,$ as $\Lambda_2(\phi)\,$ if $\,y_1<y<y_2\,$ and 
$\Lambda_3(\phi)\,$ if $\,y>y_2\,$. 

 The most general 
five-dimensional metric that respects four-dimensional 
Poincar\'{e} symmetry is
$ds^2 = e^{2\,  A(y)}\ \eta_{\mu \nu}\ dx^{\mu} dx^{\nu}
 + (dy)^2 ~.~\,$ 
Then Einstein's equations can be written as$^2$
$2\,\kappa^2\,\Lambda(\phi)= {1 \over 2}\,( {\partial W(\phi) 
\over \partial \phi} )^2
- {1 \over 3}\, W(\phi)^2 \, ,$
$
\phi' = {\partial W(\phi) \over \partial\phi} \,,
   A' = -{1 \over 6}\, W(\phi) \,,$
subject to the constraints
$W(\phi)\Big|^{y_i+\epsilon}_{y_i-\epsilon} = 
 {2\,\kappa^2} V_i(\phi_i)\,,
{\partial W(\phi) \over \partial\phi}\Big|^{y_i+\epsilon}_{y_i-\epsilon} = 
 {2\,\kappa^2}  {\partial V_i(\phi_i) \over \partial\phi}\,,$
where  $\phi_i \equiv \phi(y_i)$ and $W(\phi)$ is
 any sectionally continuous
function (which we call the superpotential), with sectional functions 
$W_{i}(\phi)$ defined analogous to $\Lambda_{i}(\phi)$. 

With this formalism in place, we turn to cosmology. Consider a metric of the 
form
$ds^{2}=-n^{2}(\tau,y) d\tau^{2}+a^{2}(\tau,y)d{\mathbf{x}^2}+
dy^{2}$,
which encodes our assumption of a static stabilizing potential. 
 We can study the contribution of matter energy densities 
on the observable brane  as a perturbation to the brane tension by making the 
ansatz
$ \rho_0=\rho+V_0 \,,\  p_0=p-V_0\,,$
where functions with the subscript $0$ are evaluated 
on the observable brane, and $\rho$ and $p$ are the perturbations. 
By requiring $e^{2A}$ to be symmetric on either side of the observable brane,  
{\it i.e.},
$W(\phi(+\epsilon))=-W(\phi(-\epsilon))\,,$
it is possible to show that
\mbox{$\left({\dot{a_0} \over a_0}\right)^2 + {\ddot{a_0} \over a_0}= 
{\kappa^4\over 36}\,V_0\,(\rho-3\,p)-{\kappa^4\over 36}\,\rho\,(\rho+3\,p)\,,$}
where we have used
$\kappa^2\,\vev{\check{T}_{55}}\,={1 \over 6}\,\vev{W(\phi_0)^2}=
{\kappa^4 \over6}\,V_0^2\,$ and $2\, \vev{\check{T}_{55}}
=\check{T}_{55}(+\epsilon)-\check{T}_{55}(-\epsilon)$. The leading term on the right-hand side reproduces the
standard cosmology if we make the identification, $\kappa^4V_0=6/M_{Pl}^2$. 
Note that a specific form of the bulk energy-momentum 
tensor was not chosen in the derivation. 
In introducing the perturbation, it is
no longer obvious that the equation of motion of $\phi$ remains consistent. 
On requiring 
that $\phi$ and $\Lambda(\phi)$ 
be unchanged before and after the introduction of the matter 
energy density, consistency requires, 
$ \left(3\, \fpa+\fpn \right)\Big|_{0\,,\,Static}=\left(3\, 
\fpa+\fpn \right)\Big|_{0\,,\,Perturbed}\ \,,$ which is the condition 
for a radiation-dominated (RD) universe, $\rho=3\,p$.
 This result is consistent with the
fact  that the radion couples to the 
trace of the energy-momentum tensor. 
It may be possible to identify the process of radion stabilization with
inflation and reheating 
and the time at which the inter-brane spacing 
becomes stable marks the end of reheating.
The RD universe then ensues. 

\begin{figure}[h]
\vspace*{13pt}
\centerline{\psfig{file=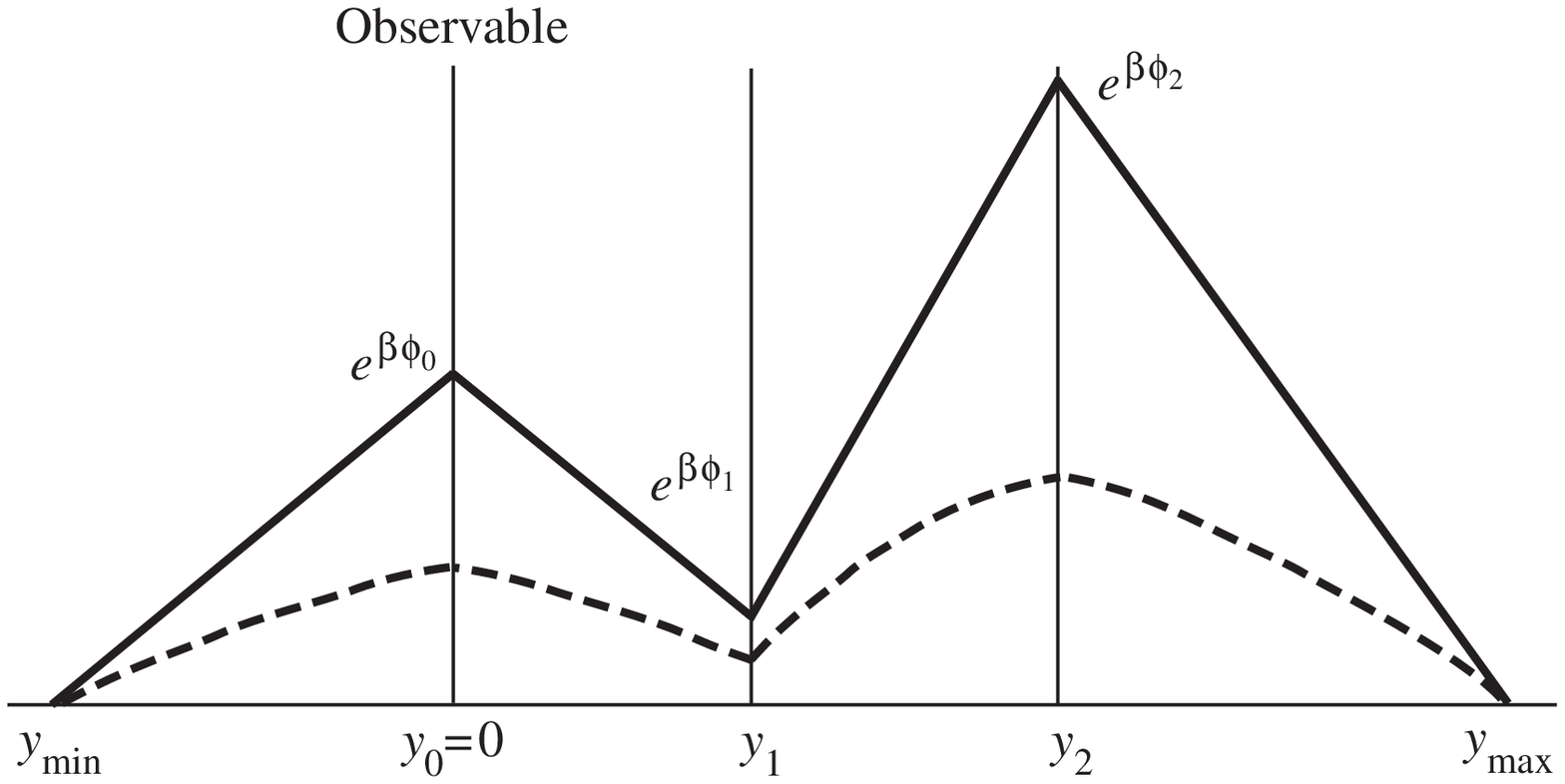,width=8cm,height=3.5cm}}
\vspace*{13pt}
\fcaption{The profile of $e^{\beta \phi}$ (solid) and  $e^{2 A}$ (dashed) in the case of self-tuning branes. 
 In regions where $\omega_i$
is positive (negative), $e^{\beta \phi}$ falls (rises) linearly.
The brane at $y_1$ has negative tension.}
\label{self}
\end{figure}

Let us address the hierarchy problem in the context of two stabilizing 
potentials that have received considerable interest.$^{3,4}$
Consider a superpotential of the exponential form 
$W_{i}(\phi)=\omega_{i}\, e^{-\beta \phi}\,,$ for which 
$12\,\kappa^2\,\Lambda_{i}(\phi)= (3\, \beta^2-2)\, 
\omega_{i}^2\, e^{-2 \beta \phi}\,.$ 
For $\beta^2=2/3$, we have the important result that $\Lambda=0$.$^3$ 
 With  $\beta^2=2/3$
 the branes are flat and will remain so, independent of the matter on 
them (hence the expression ``self-tuning flat branes''). It can be shown that
$f(y) \equiv e^{\beta \phi(y)} = -{2\over 3}\,\omega_i\,y+c_i\   {\rm{for}}\ 
 y_{i-1}\leq y \leq y_i\,,$ and $e^{2\,A(y)}=\sqrt{e^{\beta \phi}}$. Note
that when $f$ vanishes, $\phi$ diverges and $e^{2A}$ vanishes. 
  We  truncate the space at the horizons defined by  $e^{2A}=0$.
We can calculate the four-dimensional Planck scale in terms of the 
five-dimensional Planck scale,$^5$ 
$M_{Pl}^2 =  M_X^3\,\int e^{2\,A(y)}\, dy$. The result is 
$M_{Pl}^2 =  M_X^3\,\left[(
{1 \over \omega_1}-{1 \over \omega_0})\,e^{\phi_0/\beta}+
 ({1 \over \omega_2}-{1 \over \omega_1})\,e^{\phi_1/\beta}+
 ({1 \over \omega_3}-{1 \over \omega_2})\,e^{\phi_2/\beta}\right]\,.$
We recall that the electroweak scale ($M_{EW}$) 
can be generated from the five-dimensional Planck scale $M_X$
via $M_{EW} \simeq M_X\,e^{A(0)}=M_X\,e^{\phi_0/ (6\,\beta)}\,.$ 
In a two brane geometry, imposing the cosmological requirement 
that $W(\phi)^2$ be symmetric 
about the observable brane results in  $\phi$ always 
taking its maximum value on the observable brane thus making it
impossible to solve the hierarchy problem. We have shown$^1$
 that Fig.~\ref{self} displays 
the unique minimal configuration
from which the hierarchy of scales can be obtained without fine-tuning. 
There are two positive tension branes (one of which is the observable
brane), and one negative tension brane. By inspecting 
$ {12\,\kappa^2}\,V_i(\phi) = (\omega_{i+1}-\omega_{i})\, e^{-\beta \phi}\,$
it can be seen that due to the
exponential dependence of the brane tensions on $\phi\,$, 
a large hierarchy is generated between the values of the 
tensions for even moderately different values of
$\phi$.

If we consider the type of superpotential that leads to the 
stabilization mechanism suggested in Ref.$^4$, we  find that it is 
not possible to generate the appropriate scale hierarchy with only positive 
tension branes and that it is necessary for the radion to be unbounded 
for $y>y_1$. We therefore study a model with two branes where the 
hidden brane has
 negative tension. Table \ref{table} 
shows our particular choice of the polynomial superpotential
and  the solution to Einstein's equations.
 The location of the hidden brane is 
$y_1= {1 \over 2} \ln {{\phi_1 \over \phi_0}}\,,$ the
electroweak scale is 
$M_{EW} \simeq M_X\,e^{-{\phi_0^2 \over 24}}\,$ and
the Planck scale is given by
 $\left(2\,{M_{Pl} \over M_X}\right)^2 =  
\left({\phi_0^2 \over 12}\right)^{-{\xi \over 12}}\, 
\Gamma \left({\xi \over 12},{\phi_1^2 \over 12},\infty\right)
+ \left({\phi_0^2 \over 12}\right)^{-{\eta \over 12}}\,
\left[\Gamma\left({\eta \over 12},
{\phi_0^2 \over 12},{\phi_1^2 \over 12}\right)+\,
\Gamma\left({\eta \over 12},{\phi_0^2 \over 12},
{\infty}\right)\right]\,,$ where 
$\Gamma (a,x,y) \equiv \int_{x}^{y} t^{a-1}\, e^{-t}\, {dt}\,$.
Consistency conditions imposed by positivity of the tension of the
observable brane and the profile of the radion are
$\eta<\phi_0^2<\phi_1^2\,$.
When the correct hierarchy is generated, by far 
the dominant contribution to $M_{Pl}$ comes from the 
integral over the space $y>y_1$. The condition under which
this integral dominates is $\phi_0^2<\phi_1^2<\xi$. 
Then $\eta<\xi\,$, and the brane at $y_1$ has negative tension. 

For both the potentials considered,  it is not 
possible to place the negative tension brane 
at the fixed point of an orbifold because the space beyond $y_1$ is crucial 
for generating the scale hierarchy.
As a result of this, the radion may have a problem with 
positivity of energy. If one accepts this
 unpleasant circumstance the models are theoretically feasible.

\begin{table}[h]
\begin{center}
\begin{tabular}{|c|c|c|c|} 
$ W(\phi)/M_X$ & $\phi(y)$ & $A(y)$ & Region   \\
\hline
$(\eta-\phi^2) $ & $\phi_0\,e^{2\,|y|} $  & $ {1 \over 6}\,\eta\,
|y|-{\phi_0^2 \over 24}\,e^{4\,|y|} $  & $ y<0  $ \\
\hline
$-(\eta-\phi^2) $ & $\phi_0\,e^{2\,y} $  & $  {1 \over 6}\,
\eta\,y-{\phi_0^2 \over 24}\,e^{4\,y} $  
& $ 0<y<y_1     $ \\
\hline
$-(\xi-\phi^2) $ & $\phi_0\,e^{2\,y} $  & $  {1 \over 6}\,
\eta\,y_1+{1 \over 6}\,\xi\,(y-y_1)-{\phi_0^2 \over 24}\,e^{4\,y} $  
& $ y_1<y     $ \\ \hline
\end{tabular}
\caption[]{\small The solution to Einstein's equations 
in a model with a polynomial superpotential. }
\label{table}
\end{center}
\end{table}

\end{document}